\documentclass[aps,prl,twocolumn,superscriptaddress]{revtex4}
\usepackage{graphicx}
\usepackage{mathrsfs}
\usepackage{bm}
\usepackage{amsmath}
\usepackage{dcolumn}
\usepackage{epstopdf}
\usepackage{dsfont}
\usepackage{amssymb}
\usepackage{tabularx}
\usepackage{array}
\usepackage{float}
\usepackage{colordvi}

\begin{document}
	\title{Engineering Corner States from Two-Dimensional Topological Insulators}
	\author{Yafei Ren}
	\affiliation{ICQD, Hefei National Laboratory for Physical Sciences at Microscale, CAS Key Laboratory of Strongly-Coupled Quantum Matter Physics, and Department of Physics, University of Science and Technology of China, Hefei, Anhui 230026, China}
	\affiliation{Department of Physics, The University of Texas at Austin, Austin, Texas 78712, USA}
	\author{Zhenhua Qiao}
	\email[Correspondence author:~]{qiao@ustc.edu.cn}
	\affiliation{ICQD, Hefei National Laboratory for Physical Sciences at Microscale, CAS Key Laboratory of Strongly-Coupled Quantum Matter Physics, and Department of Physics, University of Science and Technology of China, Hefei, Anhui 230026, China}
	\author{Qian Niu}
	\affiliation{Department of Physics, The University of Texas at Austin, Austin, Texas 78712, USA}
	
\begin{abstract}
		We theoretically demonstrate that the second-order topological insulator with robust corner states can be realized in two-dimensional $\mathds{Z}_2$ topological insulators by applying an in-plane Zeeman field. Zeeman field breaks the time-reversal symmetry and thus destroys the $\mathds{Z}_2$ topological phase. Nevertheless, it respects some crystalline symmetries and thus can protect the higher-order topological phase. By taking the Kane-Mele model as a concrete example, we find that spin-helical edge states along zigzag boundaries are gapped out by Zeeman field whereas in-gap corner state at the intersection between two zigzag edges arises, which is independent on the field orientation. We further show that the corner states are robust against the out-of-plane Zeeman field, staggered sublattice potentials, Rashba spin-orbit coupling, and the buckling of honeycomb lattices, making them experimentally feasible. Similar behaviors can also be found in the well-known Bernevig-Hughes-Zhang model.
\end{abstract}
\date{\today}
	
\maketitle
\textit{Introduction---.} Since the discovery of $\mathds{Z}_2$ topological insulators (TIs)~\cite{Z2_KM_05,BHZ_06}, which exhibit spin-helical gapless edge modes protected by time-reversal symmetry, topological phases and materials have been extensively explored in two- and three-dimensional (3D) systems~\cite{Rev_TI_Qi_11, rev_TI_Kane_10, rev_TBandT_Bansil_16, rev_Ren_16}.
Recently, the topological phases have been generalized to higher order~\cite{Multipole_PRB_17, Multipole_Sci_17, HOTI3dTRSB_18, HOTIspinlessTRS_19, AllCase_disclination_19, CornerState_SOC_Bi_TRS_19}.
In a 3D higher-order TI, gapless 1D hinge states appear between two gapped surfaces~\cite{HOTI3dTRSB_18, HOTIspinlessTRS_19, FangChen_Hinge, Bernevig_hinge, PengY_Hinge}. Such hinge states are observed experimentally in bismuth~\cite{TCI_HOTI_Bi19, CornerState_SOC_Bi_TRS_19, Bi_3dHOTI_EXP_18}. In 2D higher-order TI, pioneering theoretical works suggest the presence of 0D corner states inside the band gap of the insulating edge and bulk~\cite{Multipole_PRB_17, Multipole_Sci_17, HOTI3dTRSB_18, TCI_HOTI_Kagome_Ezawa, PengY_Hinge}. Currently, only limited material candidates have been proposed theoretically to host 2D higher-order TI phase, including black phorsphorene~\cite{SpinlessBP_Ezawa_18}, graphyne~\cite{HOTI_Graphyne1, HOTI_Graphyne2, HOTI_Graphyne3}, and twisted bilayer graphene at certain angles~\cite{HOTI_TwistedBLG}, where the spin degrees of freedom are all neglected. Experimentally, the corner states characterizing 2D higher-order TIs have not yet been observed. As abundant candidates of $\mathds{Z}_2$ TIs have been reported~\cite{rev_Ren_16, 2DTI_Compute19, 2DTI_Compute19_2, 2DTI_Compute20, TopoMater_17, TopoMater_17_Vishwanath, TopoMater_19_Fang, TopoMater_19_Wang, TopoMater_19_Wan}, it would be of great significance to make a bridge between TIs and higher-order TIs in 2D systems.
	
\begin{figure}
	\includegraphics[width=8cm, angle=0]{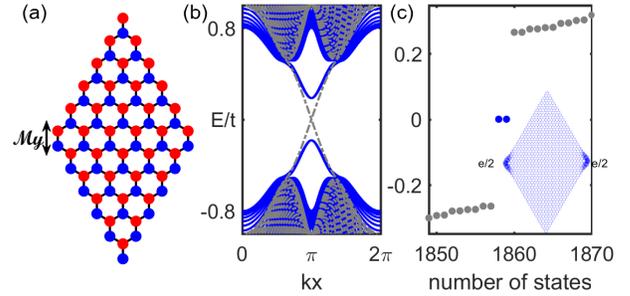}
	\caption{(a) Schematic plot of diamond-shaped honeycomb lattice nano-flake with zigzag boundaries. (b) Energy bands of the zigzag nanoribbon. The bands with and without the in-plane Zeeman field are shown in blue and gray, respectively. (c) Energy levels for diamond-shaped nanoflake. Corner states are highlighted in blue. Probability of the corner state is plotted in the inset.}\label{KMCorner}
\end{figure}
	
In this Letter, we theoretically propose to engineer higher-order TIs from 2D TIs by applying an in-plane Zeeman field. As a seminal 2D TI system, we take the Kane-Mele honeycomb lattice as a concrete example. By introducing the in-plane Zeeman field, we show that the spin-helical edge modes along the zigzag boundary become gapped, whereas robust in-gap corner states appear at the intersect between two zigzag boundaries. The fractional charge appears at each corner at half-filling. We attribute such a higher-order TI phase to the bulk topology protected by mirror-reflection symmetry $\mathcal{M}_y$, which is preserved in the presence of the in-plane Zeeman field. Along high symmetric lines $\Gamma$-K-M-K$'$-$\Gamma$ that are invariant under $\mathcal{M}_y$, we can divide the energy bands into two sub-Hilbert spaces with even and odd mirror eigenvalues. Nonzero mirror-graded winding numbers are obtained, indicating a nontrivial second-order band topology. Such bulk band topology also guarantees gapless edge modes along the armchair boundaries that respect such symmetry, manifesting the coexistence of conventional topological crystalline insulator.
As a necessary extension, we also show that a higher-order TI state can be induced by the in-plane Zeeman field in Bernevig-Hughes-Zhang model, which is another seminal 2D TI system. We further show that such corner states are robust against various perturbations due to the protection of an energy gap at the system boundaries.
	
\textit{System Model Hamiltonian---.} In our study, we first focus on the modified Kane-Mele model, i.e., honeycomb lattice in the presence of intrinsic spin-orbit coupling and in-plane Zeeman field. The corresponding tight-binding Hamiltonian can be expressed as following~\cite{Z2_KM_05}:
\begin{eqnarray}
H & =& -t\sum_{\langle ij \rangle}c^\dagger_{i}c_{j}+ i t_{\rm{SO}}\sum_{\langle\langle ij \rangle\rangle}\nu_{ij}c^\dagger_{i}{s}_{z}c_{j} \\ \nonumber
&+& \lambda \sum_{i}c^\dagger_{i} {\mathbf{B}} \cdot {\bm{{\rm s}}} c_{i} + \Delta  \sum_{i} \xi_i c^\dagger_{i}c_{i}
\label{EQ:SingleH}
\end{eqnarray}
where $c^\dagger_{i}=(c^\dagger_{i\uparrow},c^\dagger_{i\downarrow})^{{T}}$ is the creation operator for an electron with spin up/down ($\uparrow$/$\downarrow$) at the $i$-th site. The first term is the nearest-neighbor hopping with an amplitude of $t$. The second term is the intrinsic spin-orbit coupling involving next-nearest-neighbor hopping with $\nu_{ij}={\bm{d}_i \times \bm{d}_j}/{|\bm{d}_i \times \bm{d}_j|}$ where $\hat{\bm{d}}_{ij}$ is a unit vector pointing from site $j$ to $i$. The third term is Zeeman field along direction of $\mathbf{B}=(B_x,B_y,0)$ with a strength of $\lambda$, which can be introduced by either in-plane magnetic field or exchange coupling to ferromagnetic substrate with in-plane anisotropy. The last term corresponds to the staggered site energy with $\xi=\pm 1$ for different sublattices. Hereinbelow, we set $\lambda=0.2t$ and $t_{\rm SO}=0.1t$ without loss of generality, and take ${\mathbf{B}}$ along $\hat{y}$ direction and site energy $\Delta=0$ unless otherwise noted.
	
By transforming the above Hamiltonian into momentum space, one can obtain
\begin{eqnarray}
H(\bm{k})=[f_x(\bm{k})\sigma_x +f_y(\bm{k})\sigma_y]s_0 + f_{\rm SO} \sigma_z s_z + \lambda \sigma_0 s_y, \label{Hk}
\end{eqnarray}
where
\begin{eqnarray}
f_x(\bm{k})&=&t(1+2\cos {3ak_y}/{2}\cos {\sqrt{3}ak_x}/{2}), \nonumber  \\
f_y(\bm{k})&=&2t\sin {3ak_y}/{2}\cos {\sqrt{3}ak_x}/{2},  \nonumber \\ \nonumber
f_{\rm SO}(\bm{k})&=&-2t_{\rm SO}(\sin\sqrt{3}ak_x-2\cos {3ak_y}/{2}\sin {\sqrt{3}ak_x}/{2}),
\end{eqnarray}
with $\bm{k}=(k_x, k_y)$ being quasi-momentum and a being the lattice constant. $\bf{\sigma}$ and $\mathbf{s}$ are Pauli matrices for sublattice and spin, respectively.
	
\begin{figure}
	\includegraphics[width=8 cm,angle=0]{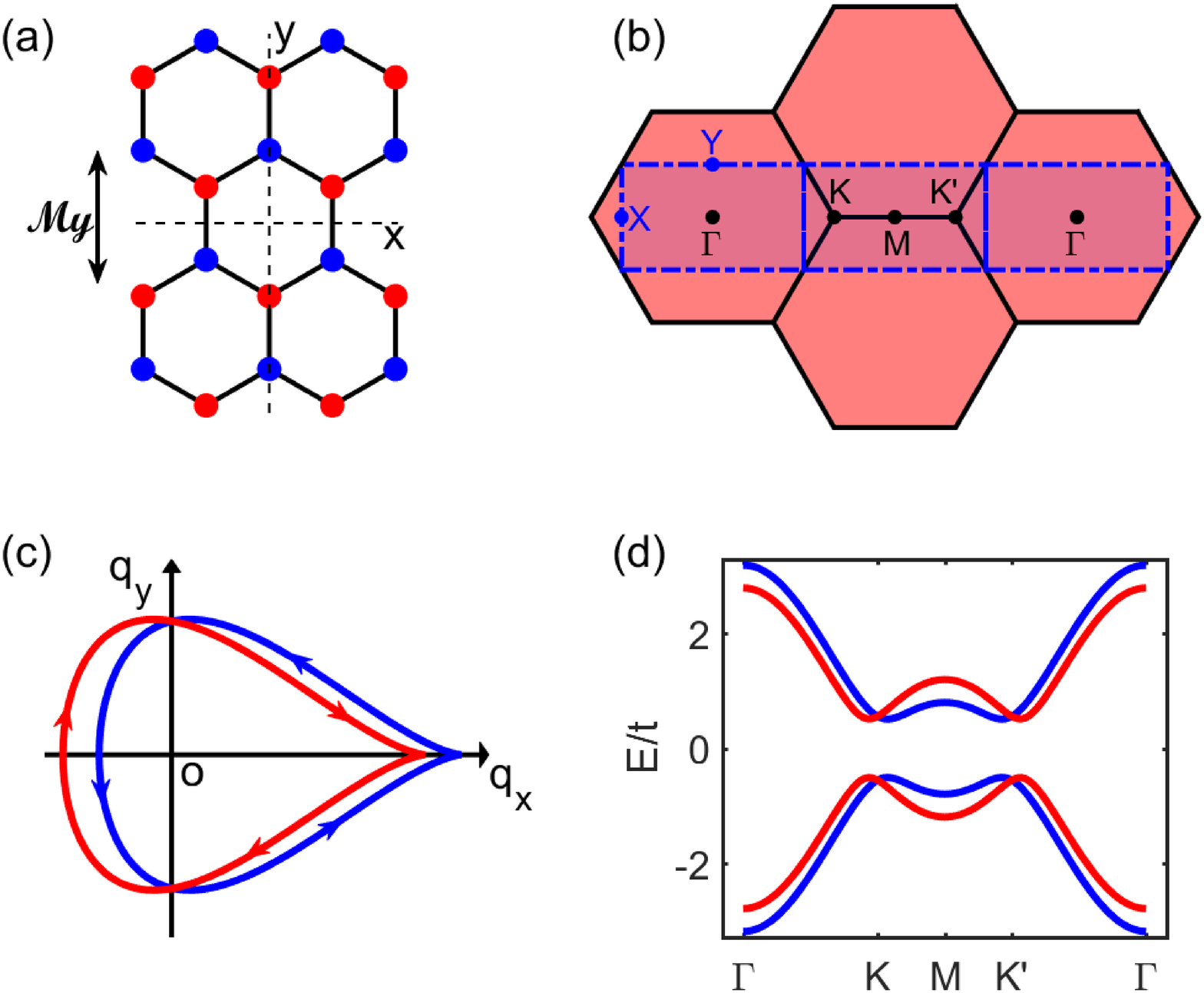}
	\caption{(a) Schematic plot of honeycomb lattices with a unit cell and supercells. (b) Brillouin zone and reduced Brillouin zone. (c) Winding of $(q_x, q_y)$ in a period. Line in red (blue) is for sub-space with $\pm i$ eigenvalue of mirror-reflection operator. The origin is enclosed for both cases suggesting the nonzero winding number for each subspace. (d) Energy bands along $k_y=0$, i.e., $\Gamma$-K-M-K$'$-$\Gamma$.}
	\label{KMband}
\end{figure}
	
\textit{Corner States---.} Here, we show the emergence of corner states. As shown in Fig.~\ref{KMCorner}(a), we first consider the corners between two zigzag-edged boundaries. In the absence of the Zeeman field, a pair of spin-helical gapless edge modes counter-propagate along the zigzag boundary as illustrated by gray lines in Fig.~\ref{KMCorner}(b) where the energy bands of a zigzag terminated nanoribbon are plotted. In the presence of the in-plane Zeeman field, the time-reversal symmetry is broken and the edge modes become gapped as shown by the energy bands in blue. Interestingly, when two gapped zigzag boundaries encounter at a corner in the diamond-shaped nano-flake, in-gap states arise as displayed in blue dots in Fig.~\ref{KMCorner}(c), where the energy levels are plotted. The probability of wavefunction at half-filling is highlighted in the inset, where we find that 1/2 electron charge is localized at each corner leading to the fractionalized charge distribution.
	
\textit{Nontrivial Bulk Band Topology---.} Now, let us move to understand the physical origin of the corner states. Although the presence of the Zeeman field breaks the time-reversal symmetry and thus drives the Kane-Mele model into a trivial insulator, an in-plane Zeeman field preserves various crystalline symmetries, e.g., inversion, mirror-reflection or rotation, making it possible to exhibit topological crystalline phases as demonstrated below. For example, if the Zeeman field is along $y$ direction, the mirror-reflection operation $\mathcal{M}_y=\sigma_x i s_y$ preserves by changing $y$ to $-y$ as illustrated in Fig.~\ref{KMband}(a), where $\sigma_x$ interchanges A/B sublattices by reflection and $i s_y$ operates on spin degree of freedom. In Fig.~\ref{KMband}(b), the high-symmetric line of Brillouin zone along $\Gamma$-K-M-K$'$-$\Gamma$ keeps invariant under $\mathcal{M}_y$ operation. Thus, $H(k_x,0)$ is invariant under the operation of $\mathcal{M}_y$.
	
By applying $\mathcal{M}_y$ twice, one can get $\mathcal{M}_y^2=-1$, meaning that $\mathcal{M}_y$ has two eigen-values of $\pm i$. The eigenvectors of $+i$ subspace are $1/2(|A \rangle \pm |B\rangle) \otimes (| \uparrow \rangle \pm i | \downarrow \rangle)$, whereas that for $-i$ subspace are $1/2(|A \rangle \pm |B\rangle) \otimes (| \uparrow \rangle \mp i | \downarrow \rangle)$, where $|A/B \rangle$ is sublattice index and $|\uparrow/\downarrow \rangle$ stands for spin up/down states. In these two subspaces, $H(k_x,0)$ can be separated into two decoupled parts:
\begin{eqnarray}
	H^{\pm}(k_x)&=&q_x^\pm \tau_x + q_y^\pm \tau_y, \\
	q_x^\pm&=&(1\pm \lambda + 2\cos {\sqrt{3}ak_x}/{2}), \\
	q_y^\pm&=&\pm 2t_{\rm SO}(\sin \sqrt{3}ak_x-2\sin {\sqrt{3}ak_x}/{2}),
\end{eqnarray}
where $\pm$ indicates the subspace with $\pm i$ eigenvalue under $\mathcal{M}_y$. One can find that both $H^{\pm}(k_x)$ exhibit chiral symmetry. Within a period of $k_x$, $q_{x,y}^\pm$ are plotted in Fig.~\ref{KMband}(c) whereas the corresponding band structures are displayed in Fig.~\ref{KMband}(d) in red and blue, respectively. In Fig.~\ref{KMband}(c), we find that $q_{x,y}^\pm$ wind around the origin (anti-)clockwisely, giving rise to mirror-graded winding numbers (also known as Zak phase) of $\nu^\pm=\pm 1$ that can be calculated by
\begin{eqnarray}
	\nu = \frac{i}{2\pi} \int {\rm d}k_x \frac{\partial_{k_x}(q_x - i q_y)}{q_x-i q_y},
\end{eqnarray}
in the system with chiral symmetry. This nonzero mirror-graded winding number indicates the system a second-order TI with in-gap corner states~\cite{Lecture}.
	
For arbitrary Zeeman field orientations in $x$-$y$ plane, e.g., $\mathbf{B}$=$(\cos \theta, \sin \theta, 0)$, one can simply perform a rotation of spin about $z$-axis to make the Zeeman field pointing along $y$-direction by applying a unitary transformation matrix $\sigma_0 \otimes \exp[-i (\pi/2-\theta)/2s_z]$. Under this transformation, the lattice and spin-orbit coupling terms in Eq.~\eqref{EQ:SingleH} remain the same. Then, the above analysis can be performed similarly. In other words, the Hamiltonian with an in-plane Zeeman field along any direction can be continuously transformed into Eq.~\eqref{EQ:SingleH}, which guarantees that the Hamiltonians with arbitrary in-plane Zeeman field are topologically equivalent.
	
\begin{figure}
	\includegraphics[width=8 cm, angle=0]{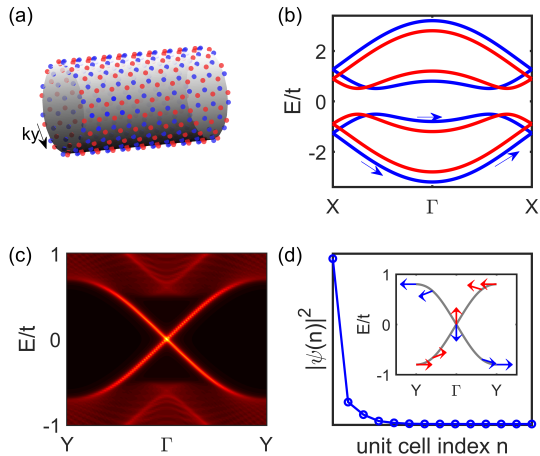}
	\caption{(a) Torus geometry of an armchair ribbon. (b) Energy bands in reduced Brillouin zone. (c) Energy bands along the left boundary of cylinder. (d) Wavefunction and expectation values of spin operator for each edge states.}
	\label{KMCylinder}
\end{figure}
	
\textit{Gapless Edge Modes in Armchair Boundaries---.} The bulk band topology discussed above also suggests the coexistence of conventional topological crystalline insulator phase as demonstrated below. Let us consider a cylinder of honeycomb lattice as illustrated in Fig.~\ref{KMCylinder}(a). In such a geometry, if one considers periodic boundary conditions along both $y$ and $x$ directions, the unit cell becomes twice larger and the Brillouin zone reduces into a rectangular-shape in blue dashed boundaries [see Fig.~\ref{KMband}(b)]. In the reduced Brillouin zone, each energy band plotted in Fig.~\ref{KMband}(c) is folded into two as shown in Fig.~\ref{KMCylinder}(b). Such a simple fold of energy band does not change the band topology, thus the band gap in Fig.~\ref{KMCylinder}(b) is also topologically nontrivial.
	
A nonzero 1D winding number of $H(k_x,k_y=0)$ corresponds to the presence of zero-energy states of a cylinder with open boundary condition along $x$ at $k_y=0$. Such in-gap states indeed appear as shown in Fig.~\ref{KMCylinder}(c), where the density of states $\rho(k_y, E)$ at the left boundary is plotted. Figure~\ref{KMCylinder}(d) plots the probability of the in-gap state and one can find that it is indeed a localized state at the boundary. When $k_y$ deviates from zero, the doubly-degenerate zero-energy states become splitting due to the coupling between both subspaces. As $k_y$ changes gradually to the Brillouin zone boundary, the degenerate zero-energy states gradually split and form gapless Dirac dispersions. Inset of Fig.~\ref{KMCylinder}(d) further exhibits the expectation value of spin operator $\bm{s}$ by red and blue arrows for edge modes with positive and negative Fermi velocities, respectively. Each arrow indicates the expectation of spin operator $\bm{s}$. One can find that the edge modes at zero-energy are fully spin-polarized along $z$ direction, and the two edge modes at the same $k_y$ display opposite spin polarization, indicating that the edge modes can be utilized to realize spin-dependent transport functions.
	
It is noteworthy that such gapless edge modes appear only along armchair boundaries where reflection symmetry preserves. When the boundary condition becomes deviated, the edge states become gapped. Such mirror-symmetry protected gapless edge modes suggests that the system is also a 2D topological crystalline insulator~\cite{TCI_Fu_11, TCI_Fu_15, TCI_Rev_16, TCI_Rev1_16}, which coexists with higher-order TI phase simultaneously~\cite{Bi_3dHOTI_EXP_18}.
	
\textit{Extension of Corner States in BHZ Model---.} We extend our findings to the other seminal 2D TI system (i.e. BHZ model). Its tight-binding Hamiltonian can be expressed as~\cite{BHZ_06}:
\begin{eqnarray}
	H(\bm{k})&=&[\frac{\epsilon_s+\epsilon_p}{2}-(t_{ss}-t_{pp})(\cos k_x + \cos k_y)] \tau_0 \otimes s_0 \nonumber \\
	&+& [\frac{\epsilon_s-\epsilon_p}{2}-(t_{ss}+t_{pp})(\cos k_x + \cos k_y)] \sigma_z \otimes s_0 \nonumber \\
	&+& 2 t_{sp} \sin k_x \sigma_x \otimes s_z + 2t_{sp}\sin k_y \sigma_y\otimes s_0,
\end{eqnarray}
where $\epsilon_{i}$ is the site energy of $i$ orbitals and $t_{ij}$ is the hopping energy between $i$ and $j$ orbitals with $i/j$ being $s$ or $p$. $\bf{\sigma}$ and $\mathbf{s}$ are Pauli matrices for orbital and spin, separately. For simplicity, we set $\epsilon_s+\epsilon_p=0$ and $t_{ss}-t_{pp}=0$, which preserve the same system topology. With in-plane magnetic field or orbital magnetization, the effective Zeeman field can be applied by $g [\alpha \sigma_z + (1-\alpha)\sigma_0] \otimes \lambda(B_x s_x+B_y s_y)$, where the difference between $g$ factors of $s$ and $p$ orbitals is dominated by the parameter $\alpha$~\cite{QAHE_InPlane_CXLiu_13}. Unless otherwise noted, we set $\alpha=0$, corresponding to a uniform $g$ factor.
	
\begin{figure}
	\includegraphics[width=8 cm, angle=0]{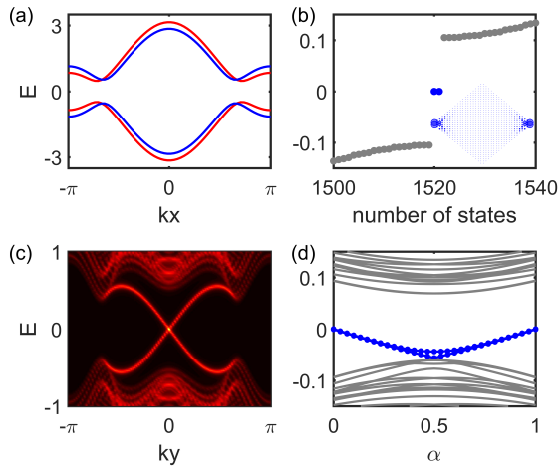}
	\caption{(a) Bulk bands along high-symmetric line $k_y=0$. Bands in even and odd sub-Hilbert spaces are shown in red and blue, separately. (b) Energy levels of a square shaped system with 760 sites. Corner states are displayed in blue. The probability of one corner state is plotted in the inset, with the circle size representing the strength. (c) Energy bands along the left cylinder boundary. (d) Energy levels of a square shaped system with 180 sites vs $\alpha$. Here, we set $\epsilon_{s}=-\epsilon_{p}=1$, $t_{ss}=t_{pp}=0.5$, $t_{sp}=0.3$, $m_x=0$, and $m_y=0.15$. $\alpha=0$ in panels (a)-(c).}
	\label{FigBHZ}
\end{figure}
	
Similar to the modified Kane-Mele model, mirror-reflection symmetry preserves along the high symmetric line of $k_y=0$, and the Hilbert space can be separated into two decoupled parts with mirror graded winding numbers. Energy bands of both parts are respectively plotted in red and blue [see Fig.~\ref{FigBHZ}(a)]. For a square-shaped system with a finite size, we also find two nearly degenerate in-gap states inside the band gap as displayed in Fig.~\ref{FigBHZ}(b). Inset plots the probability of one in-gap state where the wavefunction is equally distributed at both corners. The presence of corner states can be attributed to the same topological origin as discussed above. Thus, this BHZ model also manifests itself a topological crystalline insulator with symmetry protected gapless edge states as shown in Fig.~\ref{FigBHZ}(c), where energy bands along the left boundary of a cylinder are plotted. In the presence of nonzero $\alpha$, which means the $g$ factors of $s$ and $p$ orbitals are different, shifts the energy of corner states away from zero as displayed in Fig.~\ref{FigBHZ}(d). Nevertheless, except $\alpha=0.5$, corner states are still protected by an energy gap from the bulk states. When $\alpha=1$, $s$ and $p$ orbitals experience opposite Zeeman splittings and the in-gap states become zero-energy again, which are also corner states lying at the upper and lower corners. This indicates that anti-ferromagnetic order in the BHZ model can also induce a higher-order TI phase when both orbitals have the same $g$ factor.
	
\textit{Summary and Discussion.---} We reveal the nontrivial bulk band topology of Kane-Mele and BHZ models in the presence of the in-plane Zeeman field introduced by either magnetic field or ferromagnetic substrate. We show that the one-dimensional Hamiltonian along high-symmetric line exhibits nonzero winding number protected by mirror-reflection symmetry. Along the system boundaries respecting such reflection symmetry, gapless edge modes preserve, whereas along boundaries without such symmetry, the edge states of topological insulator phase become gapped.
Nevertheless, the higher-order TIs characterized by robust corner states appear when two gapped boundaries meet at a corner that preserves such symmetry locally. At half-filling, the fractionalized electrical charge is distributed at each corner. The bulk band topology and corner states are shown to be robust against substrate-induced extrinsic Rashba spin-orbit coupling and lattice-buckling-induced intrinsic Rashba spin-orbit coupling (see Appendix), which suggest that the above higher-order TI phase can be realized by introducing ferromagnetism with in-plane anisotropy to topological insulators. The Kane-Mele model can be applied to silicene, germanene, stanene~\cite{Si_Ham_Yao_11, QAHE_Si_Ezawa_12, TI_SiGeSn_Ezawa_15, TI_SiGeSn_SCZhang_17}, Pt$_2$HgSe$_3$, Pd$_2$HgSe$_3$~\cite{KM_Electronic_18, 2DTI_Compute19_2} whereas the BHZ model can be applied to other various topological materials~\cite{KM_Electronic_18, 2DTI_Compute19_2}. By combining the abundant 2D topological insulator materials~\cite{2DTI_Compute19, 2DTI_Compute19_2, 2DTI_Compute20} with the 2D ferromagnets as developed recently with in-plane anisotropy~\cite{CrX3_1, CrX3_PNAS_19, CrI3_MagnetismAnisotropy, CrCl3_InPlaneAFM, VSe2_NanoRes, VSe2_NatNano, NiPS3_InPlaneFM, NiPS3_PRL, Rev_2DMag, QAHE_InPlane_OsCl3_17, QAHE_InPlane_FengLiu_18}, the higher-order TI phase is highly expected to be observed.
	
Considering the application of Kane-Mele model and BHZ models in nontrivial phases of atomic crystal layers, photonic, and phononic crystals~\cite{KM_Phononic_15, KM_Phononic_19, KM_Photonic_13, KM_PhotonicRev_17, KM_Photonic_19, BHZ_Photonic, BHZ_Phononic}, our work not only reveals novel topological phases by applying in-plane Zeeman field in atomic crystals but also provides a practical way to search tunable high-order TI in spinful condensed matter systems and their bosonic analogies~\cite{HOTI_SonicCrystal_NatPhys19, Elastic_SOTI_19,  Photonic_SOTI_19, Photonic_SOTI1_19}.
	
\textit{Acknowledgments---.} The work at USTC was supported by the National Key R \& D Program (2016YFA0301700), NNSFC (11474265, 11504240), and by the Supercomputing Center of USTC and AMHPC for high-performance computing assistance. YFR was partially supported at UT Austin by the Welch Foundation (F-1255), and QN by DOE (DE-FG03-02ER45958, Division of Materials Science and Engineering) on general theoretical considerations.

\end{document}